\documentclass[12pt,preprint]{aastex}
\usepackage{emulateapj5}

\shorttitle{Possible evidence for pulsed X-rays from outer gap}
\shortauthors{Wang et al.}

\input psfig.sty
\begin{document}

\title{Possible evidence for the pulsed X-ray emission from outer gap in PSR B1937+21?}
\author{H.G. Wang, R.X. Xu and G.J. Qiao}
\affil{Department of Astronomy, Peking University, Beijing 100871,
China}

%\email{wanghg@bac.pku.edu.cn}

\begin{abstract}

The fastest millisecond pulsar PSR B1937+21 presents an interpulse
separated from the main pulse by nealy $180^{\rm o}$ at radio
frequencies. Recently, the ASCA observations (Takahashi et al.
2001) detected pulsed X-ray emission from this pulsar. Only a
single narrow X-ray pulse is observed, which is coincident with
the radio interpulse in phase. We investigate the possible origin
of the pulsed X-rays from the polar cap (PC) accelerators or the
outer gap (OG) accelerators in the frame of PC model and OG model,
respectively, by assuming a dipolar magnetic field structure and
the same radio emission pattern from its poles for the pulsar. The
OG model can naturally explain the main observational facts. For
the PC model, the coincidence between the X-ray pulse and the
radio interpulse can not be reproduced in the assumed case.
However when considering possible deviation from our assumption,
PC model may still be valid for this pulsar in some cases.

%The uncertainty of the maximum rate of position angle sweep and
%the implication of large values of this rate are discussed.

%We investigate the validity of the
%polar cap (PC) model and outer gap model (OG) for the X-ray
%emission from PSR B1937+21
%It shows that the PC model can be almost ruled out while the OG
%model can explain the main observational facts.

\end{abstract}

\keywords{pulsar: PSR B1937+21: X-rays: emission mechanism}

\section{Introduction}

Even more than 30 years after the discovering of high energy
pulsars, the theoretical reproduction of X-ray and $\gamma$-ray
emission from such pulsars is still a matter of debate. It is
commonly agreed that there are two scenarios on modelling X-ray
and $\gamma$-ray creation: the outer gap (OG) model (e.g., Cheng,
Ho \& Ruderman 1986a,b, Romani 1996, Cheng \& Zhang 1999, Hirotani
\& Shibata 2001) and the polar cap (PC) model (e.g., Harding 1981,
Sturner \& Dermer 1994, Daugherty \& Harding 1996, Luo et al.
2000, Zhang \& Harding 2000). The fundamental difference between
these two types of models is the location of the regions where
particles are accelerated to relativistic energies and emit high
energy photons. The early PC model (Harding 1981) assumed that the
emission is produced just above the PC surface. In the present
versions of PC model (e.g. Daugherty \& Harding 1996, Harding \&
Muslimov 1998), it is proposed that the particle acceleration
region may extend from the PC surface to several stellar radius
above due to free charge flow and inertial frame dragging, so that
wide double-peak X-ray and $\gamma$-ray light curves can be
reproduced, given the inclination angle is not large. In contrast,
the OG model (Cheng et al. 1986a, CHR) presumed that the gaps can
exist in the outer magnetosphere between the null charge surface
and the light cylinder. Later, CHR model was developed to the
single OG models (e.g., Chiang \& Romani 1994, Romani \&
Yadigaroglu 1995, Cheng et al. 2000), which claimed that the three
dimension extents of OGs are constrained by the pair cascade
processes, and a single OG can produce wide double-peak high
energy light curves. Whereas theoretical considerations of more
detailed physical processes for particle acceleration and photon
emission are necessary, it is urgent and interesting to find new
observational evidence for these models.

PSR B1937+21, with the period of 1.56 ms, is the fastest
millisecond pulsar (MSP) known. At radio bands, it exhibits an
interpulse emission that is roughly equal to the main pulse in
intensity and separated from it by a phase of about $180^{\rm o}$.
Recently, the ASCA observations detected pulsed X-ray emission
from this pulsar (Takahashi et al. 2001). Only one nonthermal
narrow pulse was observed, which is coincident with the radio
interpulse in phase within the timing errors. The pulse width is
about $100\pm 61$ $\mu$s ($23^{\rm o}\pm 14^{\rm o}$). Besides the
narrow pulse, the light curve reveals two additional wide
Gaussian-shaped bulges above the background level, with each phase
interval being about 0.5 rotation period and each peak intensity
$\sim1/4$ of the nonthermal pulse peak.

Where does the X-ray emission of the fastest MSP come from, the PC
or the OG? It has not been extensively studied as the other high
energy pulsars such as the Crab pulsar and Vela. The previous work
was done by Luo et al (2000), in which the PC model was modified
for MSPs. They applied the theory to PSR B1937+21 and suggested
that the X-ray emission probably originates from the location of
one stellar radius above the PC.

In this paper, efforts on modelling the observational data at
radio and X-ray bands are made for both the PC and OG models. The
inclination angle between the rotation and magnetic axes is
constrained in section 2, which is necessary for the modelling.
The modelling is carried out in section 3. Conclusions and
discussions are placed in section 4.

\section{The inclination angle}

The inclination angle ($\alpha$) between the rotation and magnetic
axes is a necessary parameter for both the PC and OG models to
give various high energy emission beams. Unfortunately, there is
no agreement on the value of inclination angle of PSR B1937+21. In
this section we reinvestigate the value of $\alpha$ under the
double-pole model, viz., the radio interpulse and main pulse are
considered to be from the opposite magnetic poles of a dipolar
field, and the result is used in the calculation in section 3.

We assume that the radio emission beams from double poles are
axisymmetric around the magnetic axis and have the same radius,
namely, $\rho_{1}=\rho_{2}$. According to the geometry model (Gil
et al. 1984,  Lyne \& Manchester 1988, LM88) one has
%
%------------ Eq. 1------------------
\begin{equation}
\sin^{2}{{\rho_{1}}\over{2}}=\sin^{2}{{W_{1}}\over{4}}\sin\alpha_{1}\sin(\alpha_{1}+\beta_{1})+\sin^{2}{{\beta_{1}}\over{2}}
\label{eq:rou1}
\end{equation}
%
%-----------  Eq. 2--------------
\begin{equation}
\sin^{2}{{\rho_{2}}\over{2}}=\sin^{2}{{W_{2}}\over{4}}\sin\alpha_{2}\sin(\alpha_{2}+\beta_{2})+\sin^{2}{{\beta_{2}}\over{2}},
%\label{eq:rou2}
\end{equation}
where $\beta$ is impact angle between the line of sight (LOS) and
the magnetic axis, $W$ is the pulse width of the average profile,
the subscripts `1' and `2' denote the main pulse and the
interpulse, respectively. There are two simple geometrical
relations between the inclination angles and the impact angles,
namely, $\alpha_{2}=\pi-\alpha_{1}$, and
$\beta_{2}=\alpha_{1}+\beta_{1}-\alpha_{2}$.

>From the above relations $\beta_{1}$ can be derived as
%
%------------- Eq. 3-----------------
\begin{equation}
\beta_{1}=\tan^{-1}\left[{{A-\tan^{2}\alpha_{1}}\over{(A+1)\tan\alpha_{1}}}\right],
\label{eq:bt1}
\end{equation}
where
%
%------------ Eq. 4----------------
\begin{equation}
A={{1}\over{\sin^{2}{{W_{1}}\over{4}}}-\sin^{2}{{W_{2}}\over{4}}}.
\end{equation}
In the following we neglect the subscript `1', so that all the
$\alpha$, $\beta$ and $\rho$ below are referred to the main pulse
except for special declaration.

To figure out the value of $A$, the profile at 1.5 GHz is chosen
(downloaded from EPN) for its high time resolution and low
dispersion smearing (Kramer et al. 1998). The pulse widths are
measured at the level of 10\% of their peak intensities, which are
$19^{\rm o}.4\pm 0^{\rm o}.4$ and $17^{\rm o}.8 \pm 0^{\rm o}.4$
for the main pulse and the interpulse respectively. So that
$\beta(\alpha)$ and $\rho(\alpha)$ can be calculated, as shown by
the solid curve and the curve dotted by circles respectively in
Fig.1. The $\beta(\alpha)$ curve approximates to a linear function
of $\alpha+\beta\thickapprox90^{\rm o}$ when
$\alpha\lesssim85^{\rm o}$, which is determined by the fact that
the main pulse and interpulse are both narrow and differ only a
little in pulse width.

For a dipolar field, the shape of the polar cap is found to change
from a circle (for $\alpha=0^{\rm o}$) to an ellipse of which the
longitudinal radius is about 1.6 times of the latitudinal radius
(for $\alpha=90^{\rm o}$, Cheng et al. 2000). Since the deviation
from circular shape is not essentially significant, we simply
regard the polar cap as a circle in this section. So that the
opening angle (between the magnetic axis and the tangent of the
magnetic field line) of the polar cap $\rho_{\rm PC}$ can be
determined by the last open field line on the plane containing the
rotation and magnetic axes (${\bf \Omega}-{\bf \mu}$ plane). The
radius $\rho_{\rm PC}$ is a function of the stellar radius R and
the inclination angle $\alpha$, as shown by the dash curves in
Fig.1 for $R=3$ km and 10 km.

We further assume that the boundary of the radio beam is defined
by the last open field lines, then the beam radius should be
greater than $\rho_{\rm PC}$. According to this criteria, the
inclination angle is constrained to be $\alpha\lesssim 63^{\rm o}$
for $R=10$ km and $\alpha\lesssim 77^{\rm o}$ for $R=3$ km. As the
radii of neutron stars (NSs) are currently believed to be about 10
km, we accept that $\alpha\lesssim 63^{\rm o}$.

Can we determine the exact value of $\alpha$? It is sure if the
maximum rate of position angle $({\rm d}\psi/{\rm d}\phi)_{\rm
max}$ is exactly known, where $\psi$ is the position angle and
$\phi$ the pulse longitude. The maximum rate presents the second
relationship between $\alpha$ and $\beta$, which reads
%
%------------- Eq.5---------------
\begin{equation}
\left({{\rm d}\psi}\over{{\rm d}\phi}\right)_{\rm
max}={{\sin\alpha}\over{\sin\beta}}  . \label{eq:ps}
\end{equation}
Combining with Eq.s~\ref{eq:bt1} and ~\ref{eq:ps}, $\alpha$ and
$\beta$ can be solved for a given $({\rm d}\psi/{\rm d}\phi)_{\rm
max}$.

Although recent polarization observations present flat position
angle sweeps (Thorsett \& Stinebring 1990, Stairs et al. 1999), it
can not be asserted that the real value of $({\rm d}\psi/{\rm
d}\phi)_{\rm max}$ is small, because observations may give a less
steep position angle gradient due to smearing of finite sampling
time, to the frequency dispersion in pulse arrival time (Liu \&
Wu, 1999), and to the interstellar scattering (Gil 1985a). The
$\beta^{\prime}(\alpha)$ curves derived from Eq.~\ref{eq:ps} for
$({\rm d}\psi/{\rm d}\phi)_{\rm max}= 1, 3$ and 20 respectively
are presented by the dotted curves in Fig.1. The intersections of
$\beta^{\prime}(\alpha)$ and $\beta(\alpha)$ shows that a larger
$({\rm d}\psi/{\rm d}\phi)_{\rm max}$ results in a larger $\alpha$
and smaller $\beta$, which means the LOS sweeps across the radio
beam more closely to the beam center.

One may find that when the real value of $({\rm d}\psi/{\rm
d}\phi)_{\rm max}$ is large enough, $\alpha$ would exceed $63^{\rm
o}$ (for example, taken $({\rm d}\psi/{\rm d}\phi)_{\rm max}=3$ as
proposed by Gil (1985a), $\alpha$ is $71^{\rm o}$) and hence
contradict against $\alpha\lesssim 63^{\rm o}$. However, if the
radius is smaller, for example, $R=3$ km, this inconsistency would
cancel. In fact, Xu et al. (2001) suggested that PSR B1937+21 is
probably a strange star (SS) with low mass and small radius. The
detailed discussion is placed in section 4.

The range of $\alpha$ presented above is different from the
conventional consideration in double-pole model that $\alpha$
should be close to 90$^{\rm o}$ (Stairs et al. 1999).
Alternatively, there is another kind of so-called single pole
model to interpret the interpulse, which suggests that the
interpulse emission comes from the same pole as the main pulse. In
the single pole model proposed by Gil (1985a) for PSR B1937+21,
$\alpha$ only need to be 20$^{\rm o}$. The single pole model
predicts that the separation between the main pulse and interpulse
may be close to 180$^{\rm o}$ and is frequency independent (Gil
1983, 1985b), which is coincident with the observation of PSR
B1937+21 (Hankins \& Fowler 1986). However, observations with high
time resolution (e.g. Kramer et al. 1998, Stairs et al. 1999)
failed to find the emission components between the main pulse and
interpulse, which was reported by Stinebring et al. (1984) and was
suggested to be a strong support to the single pole model (Gil
1985a). Therefore, in this paper the radio emission is considered
to come from double poles. More confirmative estimations of $({\rm
d}\psi/{\rm d}\phi)_{\rm max}$ are expected to determine $\alpha$
and $\beta$.

\section{The origin of nonthermal X-ray emission from PSR B1937+21}

In this section we calculate the X-ray beams in the frames of both
the PC and OG models to find out whether they are able to
reproduce the observational facts. The facts used here are:

(1) at 1.4 GHz the separation between the peaks of interpulse and
main pulse is $174^{\rm o}$ (measured from the profile presented
by Takahashi et al. (2001));

(2) at 1.5 GHz (EPN data) the 10\% widths of the main pulse and
interpulse are $19.4^{\rm o}$ and $17.7^{\rm o}$, respectively;

(3) the nonthermal X-ray pulse is nearly coincident with the
interpulse, the X-ray pulse width is about $23^{\rm o}$ (Takahashi
et al. 2001).

\subsection{Origin from the PC accelerators?}

Luo et al. (2000) discussed the viability of PC models for high
energy emission from MSPs. They found that the maximum Lorenz
factor of particles is limited by curvature radiation and not
sensitive to the specific acceleration model, but the height where
the Lorenz factor achieves the maximum is model dependent, which
may be between $0.01R$ (for the inner vacuum gap) and above $0.1R$
(for the space-charge limited gap) from PC surface for pulsar
period $P=2$ ms and a surface magnetic field $B_{\rm
s}=7.5\times10^{8}$ G. Assuming a space-charge limited flow, the
pair cascades can occur at the typical distance (to the star
center) of $r\simeq (1.5-2.5)R$, and high energy emission is
radiated from this region. With respect to $P=1.56$ ms and $B_{\rm
s}=4.1\times10^{8}$ G for PSR B1937+21, their analysis applies to
this pulsar.

Since the radio emission is also radiated from the region near the
PC, homocentric radio and X-ray beams are produced (Fig.2a).
Quantitatively, we assume the radio emission arises from the PC up
to the distance $r=2.5R$, the X-rays may arise from the PC to a
(a) relatively higher distance, e.g., $r=3R$, or (b) lower
distance, e.g., $r=2R$. For simplicity we assume the emission
regions are bounded by the last open field lines. In case (a) the
X-ray beams are wider than the radio beams, therefore, when the
line of sight sweeps across the both radio beams (to reproduce the
main pulse and the interpulse) it would sweep across the both
X-ray beams either, and hence gives double X-ray pulse, which is
inconsistent with the observation. In case (b) the X-ray beams are
narrower than the radio beams, and then, for a proper viewing
angle (between LOS and the rotation axis), only one X-ray beam
could be observed. In this case, could the fact (3) be explained
as well? We analyze this issue by using the figure of the
(phase-viewing angle) plane on which the emission beams are
projected.

In Fig.3, the horizon axis is pulse phase, the vertical axis is is
the viewing angle $\zeta$ ($\zeta=\alpha+\beta$). The dotted
curves represent the boundaries of X-ray beams, the solid curves
are the boundaries of radio beams, and the PCs are shown by the
dashed curves. Retardation (due to distinct emission heights) and
abberation effects (due to emission sources co-rotating with the
pulsar) have been taken into account, both of which make the lower
beam move towards the trailing edge of the higher beam. To obtain
the figure, a moderate inclination angle is taken, $\alpha=60^{\rm
o}$. Although there is $\alpha\lesssim 63^{\rm o}$ as discussed in
section 2, the angle should not be too small, or it would give too
wide radio beams (Fig.1), which must be emitted from unbelievably
high distances near the light cylinder.

According to Fig.3, when $\zeta\simeq80^{\rm o}$ the X-ray peak is
coincident with the radio pulse centered on phase about 0.9; when
$\zeta\simeq100^{\rm o}$ it is coincident with the radio pulse on
phase about 0.4.
However in both cases,  the radio pulse associated with the X-ray
pulse is the main pulse which is wider and more intense than the
other one.

However, we should point out that the above analysis is based on a
simple assumption that the radio (X-ray) beams from both poles
have the same width and the emission pattern in the beams is
symmetrical about the magnetic axis. Since the magnetosphere
structure of pulsars and the detailed radio emission process are
still unknown exactly, possibly significant deviation from the
assumption can not be ruled out, and the validity of the PC model
need to be considered further. We discuss such possibility in
section 4.

\subsection{Origin from the OG accelerators}

In the original OG model CHR suggested that a global current flow
through the magnetosphere can result in large regions of OGs
between the null charge surface and the light cylinder along the
last open field lines. Within the OGs particles with one kind of
charge are accelerated outward from the star and give an outward
emission beam, while those with the opposite charge are
accelerated towards the star and give an inward beam. The high
energy photons were proposed to be emitted from two OGs associated
with the two poles so that double-peaked $\gamma$-ray pulse
profile can be reproduced, of which one peak corresponds to the
outward beam from one OG and the other peak to the inward beam
from the opposite OG. CHR assumed that the OGs are active only
near the ${\bf \Omega}-{\bf \mu}$ plane. However, this assumption
is merely valid for large inclination angles.

The currently prevalent OG models are the single OG models (e.g.
Chiang \& Romani 1992, Chiang \& Romani 1994, Cheng et al. 2000).
Generally the inward emission is not important in these models for
the reason that the inward high energy photons can not pass freely
through the inner magnetosphere due to magnetic pair production.
The outward emission from the OG associated with a single pole can
produce a broad, irregularly-shaped emission beam which is
particularly dense near the edge.
The OG regions can be supported along all the last open field
lines, but the three dimension scales of OGs are limited by the
pair production processes.

In the latest version of this type of model (Cheng et al. 2000),
the fraction size ($f\equiv h/R_{\rm L}$) of the gap is $f\simeq
5.5P^{26/21}B_{12}^{-4/7}\xi^{-1/7}$, which can be estimated by
the threshold of $\gamma-\gamma$ pair production, $E_{\rm
X}(f)E_{\gamma}(f)\geqslant(m_{\rm e}c^{2})^{2}$, where $h$ is the
mean vertical extension perpendicular to the magnetic field,
$R_{\rm L}$ is the radius of light cylinder,
$\xi=\Delta\phi/2\pi$, $\Delta\phi$ is the transverse extension of
the gap, $E_{\rm x}$ is the energy of the X-ray photons emitted
from hot PCs, and $E_{\gamma}$ is the characteristic photon energy
emitted by the relativistic particles. The radial scale of pair
production is limited to a range between $r_{\rm in}$ and $r_{\rm
lim}\sim 6r_{\rm in}(\xi=0)$, where $r_{\rm in}$ (the subscript
`in' means the inner boundary of the OG) is the distance of null
charge surface, $\xi=0$ corresponds to the ${\bf \Omega}-{\bf
\mu}$ plane.

In the following our modelling is in the frame of single OG model.
Only the outward emission beams from two OGs are considered, as
illustrated by Fig.2b. For PSR B1937+21, we have a thin OG with
$f=0.16$ $\xi^{-1/7}$, so that the X-rays can be simply regarded
as being radiated from the last open field lines unless $\xi$ is
too small. The radial scale $r_{\rm lim}/r_{\rm in}$ and the
transverse scale $\xi$ are free parameters in calculating the
X-ray beams.

First we consider a general situation of the OG scenario to test
if the observation facts can be hopefully reproduced. From Fig.2b
one can see that provided the observer's viewing angle is not just
$90^{\rm o}$, e.g., $\zeta=83^{\rm o}$, the LOS can sweep across
both of the radio beams and only one X-ray beam. By assuming a
group of reasonable values of parameters (see the first line of
Table 1), namely, the inclination angle, the stellar radius, the
distance of radio emission, and the radial and transverse
extensions of the OGs, the X-ray beams are calculated, which are
demonstrated by the line-shadowed areas in Fig.3. Retardation and
abberation effects are also included. It shows clearly that the
X-ray pulse could be associated with the radio interpulse (which
has a smaller width than the other one), and the narrow X-ray
pulse width may be obtained if proper extension of the OG is
assumed.

Then we model the observational data. The parameters listed in the
second line of Table 1 are found out to be able to reproduce the
narrow X-ray pulse width, the radio pulse widths, and the
coincidence between the X-ray pulse and the radio interpulse,
which are in good agreement with the observational data. In order
to simulate the observational profiles, we simply assume that the
X-ray and radio pulses are Gaussian shapes, and assume
additionally wide, weak, hot X-ray emission from both of the PCs.
The theoretical profiles are plotted in Fig.4, together with the
observational profiles for comparison.

It should be pointed out that modelling the X-ray pulse width is
not sensitive to the value of $r_{\rm lim}/r_{\rm in}$ but to
$\xi$, thereby the range of $r_{\rm lim}/r_{\rm in}$ is relaxed
and a reasonable value is chosen. Other groups of values are also
tried. It is found that for the moderate inclination angles
$40^{\rm o}\lesssim\alpha\lesssim63^{\rm o}$ the observational
data can be reproduced with proper choice of the gap size
($\xi=40^{\rm o}$ in Table 1 is approximately the maximum
transverse scale). Therefore, our calculation suggests that the
nonthermal X-rays of PSR B1937+21 may be emitted from the OGs.
%and the pulsar may be an oblique rotator with a moderate inclination
%angle.

\section{Conclusions and Discussions}

The discovery of pulsed X-ray emission from the MSP PSR B1937+21
and the phase alignment between the X-ray pulse and the radio
interpulse provide valuable information on the emission mechanism.
In this paper we investigate the possible origin of the X-rays
from both the PC and OG accelerators. In the frame of the
prevailing OG model (e.g. Cheng et al. 2000), the X-rays from
outer gap accelerators could account for the main observational
facts by assuming proper size of the OGs for an oblique rotator.
In the frame of PC model, by assuming symmetric geometry for radio
and X-ray emission, the X-rays from polar cap accelerators would
produce an X-ray peak aligned with the main radio pulse in phase,
which is contradictive against the observation.

Some more discussions for both the OG and PC models are presented
as follows.
First we refer to the OG model.

There is a slight inconsistency as shown in Fig.4, i.e., in the OG
model, the calculated separation between the interpulse and main
pulse is $180^{\rm o}$, $6^{\rm o}$ greater than the observational
value. This may be due to the retardation effect. The phase shift
for different heights can be roughly estimated as $\Delta s=\Delta
r/(Pc)$. A difference of $\Delta r=0.8R$ is enough to produce the
phase shift of $6^{\rm o}$.

A moderate inclination angle $55^{\rm o}$ is used in the
modelling. But $\alpha$ could be larger if the real gradient of
position angle is steeper than the present observations as
suggested in section 2. Could the OG model be still valid for
large value of $\alpha$?

As shown in Fig.1, when $({\rm d}\psi/{\rm d}\phi)_{\rm max} \ga
4$ it would result in a puzzling problem that the derived radio
beam radius is considerably smaller than the PC radius, if $R=10$
km and a magnetic dipole are assumed. A much smaller stellar
radius could cancel the problem, but this requires an SS scenario,
because the smallest radius of NSs is generally believed to be
$\sim$9 km while SSs can have much smaller radius due to their
different equations of state. In fact, according to the
observational limits on the radius and mass derived from the pulse
width and $({\rm d}\psi/{\rm d}\phi)_{\rm max}$, Xu et al. (2001)
suggested that PSR B1937+21 is probably a strange star with much
low mass, small radius and weak magnetic moment. If PSR B1937+21
is an SS with small radius, for example, $R=2$ km, a group of
parameters is found out to be able to model the observational data
by OG model, which is listed in the third line in Table 1.

Then we turn to the PC model. When the emission geometry is not
symmetrical, the PC model could be able to explain the observation
as well. The possible asymmetry is discussed for the radio and the
X-ray emission, respectively.

(1) It has been assumed that the radio emission pattern from the
two poles is the same in our modelling; the different behaviors of
the main pulse and the interpulse are thus geometrical origin.
However, their difference may be caused intrinsically since the
mechanism for the radio emission is still poorly understood. For
example, a possible reason may be that the pulsar has a
non-dipolar magnetic field, thus the distribution of emission
intensity in the radio beams may be different from each other,
which could lead to that for one beam only part of it is observed
while for the other a larger part or the whole is observed.

(2) When PSR B1937+21 is an NS or an SS with crust and its
inclination angle is near $90^{\rm o}$, the X-ray emission pattern
may be significantly asymmetric about the magnetic axis. In this
case, the space charge is negative on the side toward the rotation
axis (where ${\bf \Omega\cdot B}>0$, hereafter side I) and
positive on the side away from the axis (where ${\bf \Omega\cdot
B}<0$, hereafter side II). On side II the ions can be pulled away
from the surface by strong electric field if the binding energy is
small enough. Therefore, the half beam from side II could be much
less luminous than that from side I due to the much smaller
Lorentz factors of the ions, and thus may be too weak to be
observed. Notice that, if side I is above the equator on one pole,
it should be below the equator on the other pole.
Therefore, only single X-ray peak is observed which may coincide
with the radio interpulse.

In recent years, it is suggested that pulsars may be bare strange
stars (BSSs) (e.g., Xu 2002 and references therein). If PSR
B1937+21 is also a BSS, the positive charge on side II is carried
by positrons but not ions, then the emission pattern on both sides
should be the same, and each beam may be symmetric around the
magnetic axis. In this case, when symmetrical geometry is assumed
for the radio emission, PC model could not account for the
observational facts, otherwise, PC model may still be valid.

In general, further research on the pulsar's magnetosphere
structure and emission mechanisms will be helpful to understand
the origin of its X-rays for PSR B1937+21. We are expecting that
future polarization observations could provide more confirmative
value of $({\rm d}\psi/{\rm d}\phi)_{\rm max}$, which is
meaningful not only for constraining whether this pulsar is an NS
or an SS but also for a better understanding of its X-ray
emission.

\vspace{0.2cm} %
\noindent {\it Acknowledgments}:

We are grateful to Prof. J.A. Gil and Dr. B. Zhang for their
helpful comments and discussions. The valuable suggestions from an
anonymous referee are also sincerely acknowledged. This work is
supported by National Nature Science Foundation of China
(10173002), and by the Special Funds for Major State Basic
Research Projects of China.

%\clearpage

\begin{deluxetable}{cccccccc}
\tablecaption{Parameters for calculating the radio and X-ray
beams/profiles}%
\tablewidth{0pt}%
\tablehead{ \colhead{No.} & \colhead{$\alpha$ ($^{\rm o}$)} &
\colhead{$\zeta$ ($^{\rm o}$)} & \colhead{$R$ (km)} &
\colhead{$r\tablenotemark{a}$ $$ ($R$)} & \colhead{$r_{\rm
in}(0)$\tablenotemark{b} $$ ($R$)}& \colhead{$r_{\rm lim}/r_{\rm
in}$} &
\colhead{$\Delta\phi$ ($^{\rm o}$)} }%

\startdata
1 & 60 & - & 10 & 2.5 & 1.3 & 3.5 & 100 \\
2 & 55 & 89.5 & 10 & 1.7 & 1.5 & 2.0 & 40 \\
3 & 75 & 89.5 & 2 & 2.9 & 2.1 & 2.5 & 66 \\

\enddata
\tablenotetext{a}{the distance of radio emission, in unit of
stellar radius $R$.}
%\tablecomments{b}
\tablenotetext{b}{the distance of the null charge surface on the
${\bf \Omega}-{\bf \mu}$ plane.}
\end{deluxetable}

\clearpage

%Figure 1 ---------------------------------------
\begin{figure}

\centerline{\psfig{figure=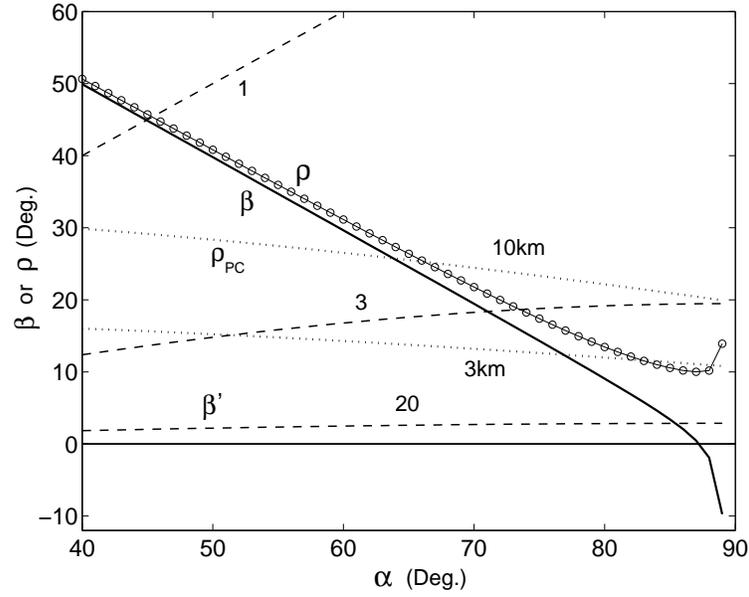,height=80mm,width=100mm,angle=0}}

\caption{Plot of $\beta$ and $\rho$ as functions of inclination
angle $\alpha$. The curves of $\beta(\alpha)$ and $\rho(\alpha)$
are derived from the observational pulse widths. The dashed curves
are $\beta^{\prime}(\alpha)$ derived from Eq.~\ref{eq:ps}, given
$({\rm d}\psi/{\rm d}\phi)_{\rm max}=1, 3$ and $20$, respectively.
The dotted curves are the opening angle of the polar cap
($\rho_{\rm PC}$), given $R=3$ km and $10$ km.\label{Figure1}}
\end{figure}

%\clearpage

%Figure 2 ---------------------------------------
\begin{figure}

\centerline{\psfig{figure=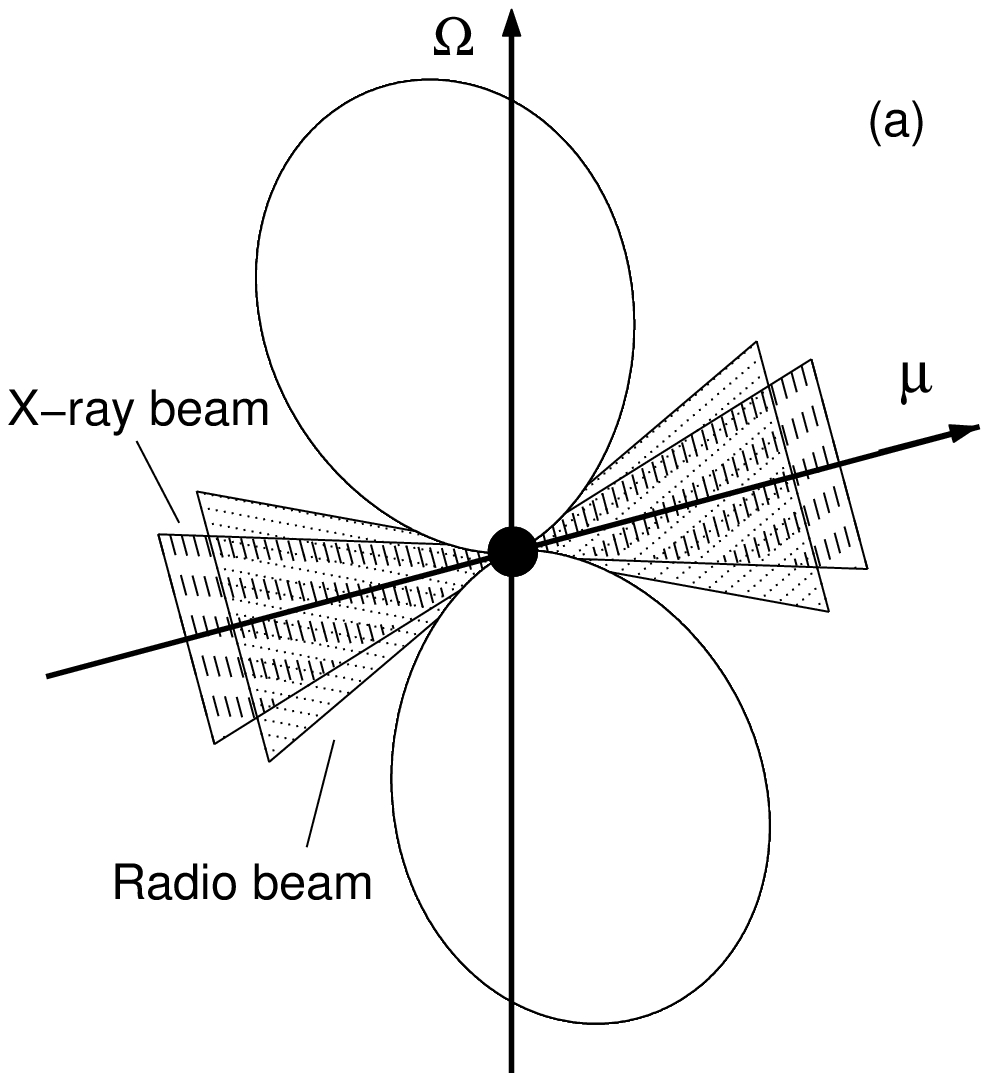,height=80mm,width=80mm,angle=0}
\psfig{figure=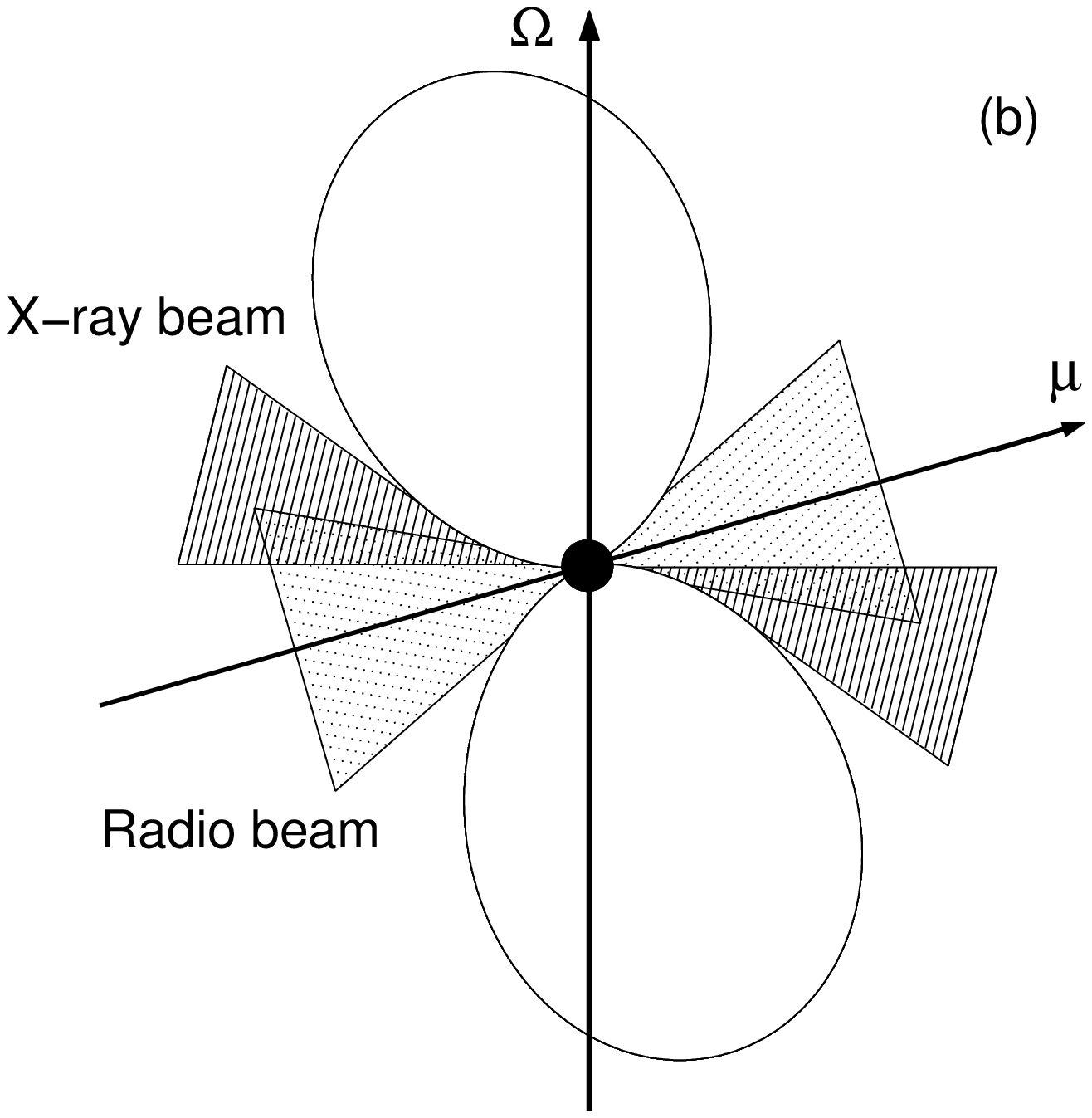,height=80mm,width=80mm,angle=0}}

\caption{(a) Scheme for the X-ray beams in the frame of PC model.
(b) Scheme for the X-ray beams produced by the OG model. Radio
beams are also plotted. \label{Figure2}}
\end{figure}

\clearpage

%Figure 3 ---------------------------------------
\begin{figure}

\centerline{\psfig{figure=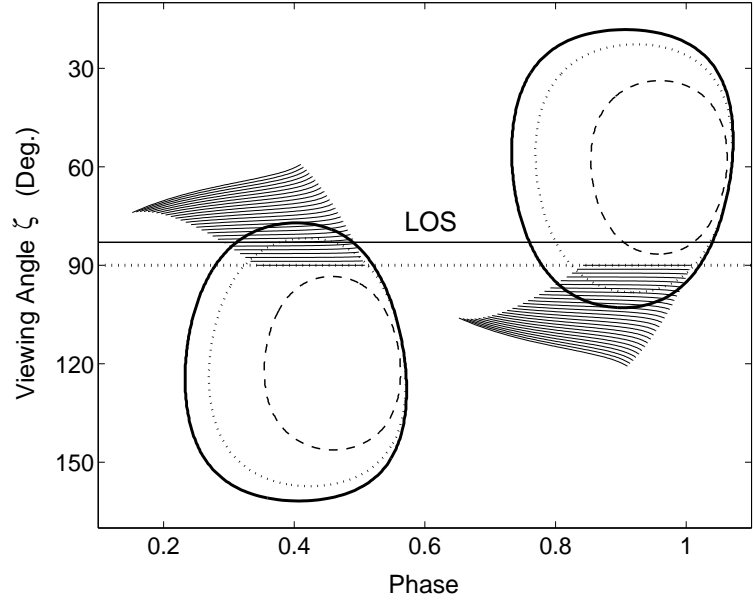,height=80mm,width=100mm,angle=0}}

\caption{Emission beams projected onto the (phase-viewing angle)
plane for $\alpha=60^{\rm o}$. The solid curves are the boundaries
of radio beams, the dotted curves are the boundaries of X-ray
beams from the extended acceleration zones above the PCs suggested
by the PC model, the line-shadowed areas are the X-ray beams from
the outer gaps, and the dashed curves represent the PCs (See text
and the first line of Table 1 for the related parameters).
\label{Figure3}}
\end{figure}

%\clearpage

%Figure 4 ---------------------------------------
\begin{figure}

\centerline{\psfig{figure=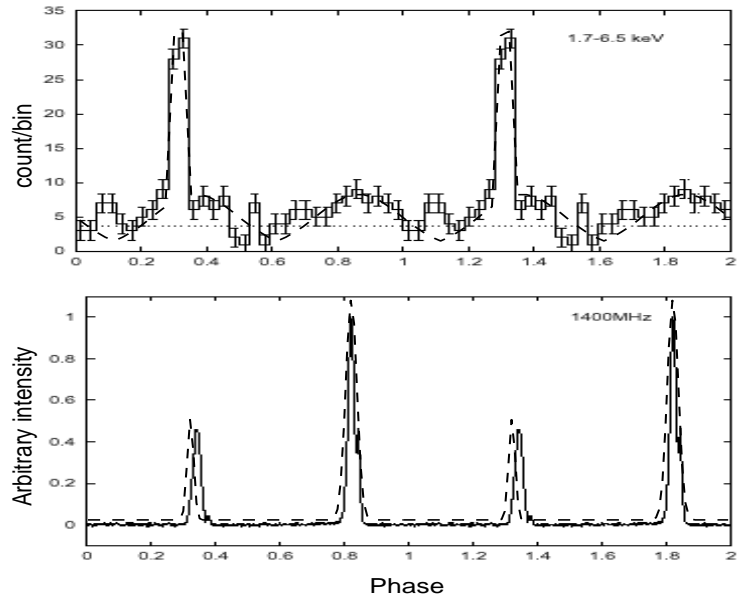,height=80mm,width=100mm,angle=0}}

\caption{Theoretical X-ray and radio profiles (the dashed curves)
together with the observational profiles (Takahashi et al. 2001).
The parameters in the second line of Table 1 are used to obtain
the theoretical profiles. \label{Figure4}}
\end{figure}

\end{document}